\documentclass[superscriptaddress,amsmath,amssymb,aps,twocolumn,prx,notitlepage]{revtex4-2} \usepackage{amsmath,amsfonts,amssymb,amsthm,graphics,graphicx,epsfig,bbm} 
\usepackage[colorlinks=true,citecolor=blue,linkcolor=blue,urlcolor=magenta]{hyperref} \usepackage[usenames]{color} \usepackage{graphicx} \usepackage{subfigure} \usepackage{amsmath} 
\usepackage{orcidlink}
\usepackage{epsfig} \usepackage{dcolumn} \usepackage{bm} \usepackage{color} \usepackage{times} \usepackage{epstopdf} \usepackage{amssymb} \usepackage{amstext} \usepackage{latexsym} \usepackage{hyperref} \usepackage{amsfonts} \usepackage{psfrag} \usepackage{soul,xcolor} \usepackage[normalem]{ulem} \usepackage{adjustbox}   \newcommand{\ket}[1]{\vert #1 \rangle}         \newcommand{\Tr}{\mathrm{Tr}}

\begin{document}
\title{Inertial Counterdiabatic Driving}

\author{Juli\'an Ferreiro-V\'elez$^{\orcidlink{0000-0003-4864-7623}}$}
\email{julian.ferreiro@tecnalia.com}
\affiliation{TECNALIA, Basque Research and Technology Alliance (BRTA), 48160 Derio, Spain}
\affiliation{Department of Physics and EHU Quantum Center, University of the Basque Country UPV/EHU, P.O. Box 644, 48080 Bilbao, Spain }

\author{F. A. C\'ardenas-L\'opez$^{\orcidlink{0000-0002-2916-2826}}$}
\email{f.cardenas.lopez@fz-juelich.de}
\affiliation{Forschungszentrum J\"ulich, Institute of Quantum Control (PGI-8), D-52425 J\"ulich, Germany} 

\author{F. Motzoi$^{\orcidlink{0000-0003-4756-5976}}$}
\affiliation{Forschungszentrum J\"ulich, Institute of Quantum Control (PGI-8), D-52425 J\"ulich, Germany} 
\affiliation{Institute for Theoretical Physics, University of Cologne, 50937 Cologne, Germany}

\author{Xi Chen$^{\orcidlink{0000-0003-4221-4288}}$}
\email{xi.chen@csic.es}
\affiliation{Quantum Advanced Research Center (QuARC), CSIC, 28049, Madrid, Spain}
\affiliation{Instituto de Ciencia de Materiales de Madrid (ICMM), CSIC, 28049, Madrid, Spain}

\begin{abstract}
We introduce an inertial counterdiabatic (CD) protocol that reformulates shortcut-to-adiabaticity design as a moving-frame diagonalization problem. The method employs the Magnus expansion to construct a variational propagator generated by an accessible CD ansatz. In the new rotating frame, the controls are optimized to suppress the off-diagonal Hamiltonian terms, and the CD terms are defined to counteract the corresponding inertial contribution. The propagator generated with the Magnus expansion naturally exploits additional control directions generated through its nested commutator, while the sequential design of the propagator introduces temporal nonlocality into the optimization. We benchmark the method on a two-qubit system and on the one-dimensional quantum Ising chain, where it consistently outperforms variational CD protocols based on action minimization. The largest improvements are observed near gap-closing regions, demonstrating the advantage of combining temporal and algebraic nonlocality for suppressing diabatic excitations. 

\end{abstract}
\maketitle

 \section{Introduction} \label{sec1}

Coherent control of quantum systems is a fundamental requirement for quantum technologies, ranging from quantum computation to quantum sensing \cite{glaser2015training,koch2022quantum,Duncanprxquantum}. Adiabatic protocols offer a natural route to high-fidelity state preparation, but the adiabatic theorem constrains their speed to scale inversely with the relevant energy gaps, making evolution times often comparable to device coherence times. Shortcuts to Adiabaticity (STA)~\cite{STA_review_RMP, focus_STA} circumvent this constraint by reproducing adiabatic outcomes on shorter timescales. Among them, counterdiabatic (CD) driving \cite{Demirplak2005,unanyan1997laser,Berry_2009,Chenprl105} introduces an auxiliary term generated by the adiabatic gauge potential (AGP) \cite{jarzynski2013} that exactly cancels nonadiabatic transitions at every instant. However, the exact AGP generally requires full spectral knowledge and involves highly nonlocal many-body interactions \cite{Adolfo_QPT_ISING_PRL}. Variational constructions \cite{sels_polkovnikov_minimizing} leverage this by approximating the AGP within a restricted operator manifold via a nested-commutator or Krylov expansion~\cite{krylov_STA,wbbs-s8fs}, retaining only few-body operators at the cost of approximate diabatic suppression.

A defining feature of standard variational AGP constructions is their temporal locality: the variational coefficients are determined pointwise along the control path from the instantaneous Hamiltonian, without explicitly incorporating the accumulated effect of the preceding dynamics. Other control strategies, such as superadiabatic expansions \cite{Superadiabatic_1,superadiabatic_2,Superadiabatic_3} and DRAG protocols \cite{Drag_1,Drag_2,Drag_3}, in contrast, account for the accumulated effect of errors, reducing them through consecutive frame transformations and destructive interference over the full evolution, respectively. However, these methods emphasize dynamical error suppression rather than instantaneous approximation of the AGP, although their construction, generalization, and fundaments differ substantially from variational CD.

The Magnus expansion (ME) \cite{magnus_og, blanes_magnus, magnus_dalgaard_integrator} provides a natural framework for retaining the time-ordered structure in driven systems. It has been applied both to suppress accumulated errors in leakage-control schemes \cite{magnuswarren1984,magnus_control_theis, magnus_control_clerk,magnus_expansion_kirchhoff} and to engineer effective CD Hamiltonians through rapidly modulated control within Floquet control \cite{claeys_floq,Floquet_CD_2,magnus_sliced}. Nevertheless, instantaneous AGP optimization and propagator-level error suppression have largely remained conceptually separate.

In this work, we bridge these viewpoints by introducing an \emph{inertial CD} protocol that employs the ME as a constructive tool for the moving-frame propagator itself. Auxiliary control fields are optimized to render the Hamiltonian as diagonal as possible within a variational moving frame built stepwise through a sequential Magnus propagator. This formulation requires no spectral information and targets off-diagonal suppression directly, without using proximity to the exact AGP as the optimization objective. This marks a fundamental distinction from standard variational AGP approaches: the optimization depends explicitly on the accumulated moving-frame propagator rather than solely on instantaneous Hamiltonian information. The conceptual distinction between the two approaches is summarized in Fig. \ref{fig: main}.  At the same time, the nested-commutator structure of the ME naturally generates higher-order operators from the dynamical Lie algebra (DLA) of the control ansatz \cite{dAlessandro, petiziol,petiziolEPL}, providing access to an enlarged effective operator space without explicitly enlarging the physical control pool.
As demonstrated below for spin systems, the accumulated moving-frame construction improves diabatic suppression in fast-driving and small-gap regimes, providing a competitive alternative to action-based variational CD within the same set of control operators. Beyond first order, by including the effective DLA directions generated by higher-order Magnus terms, we can further enhance the suppression of diabatic errors and notably reduce transitions in challenging small-gap regions, such as the finite-size critical region of the quantum Ising model.

The paper is organized as follows. In Sec. \ref{APPCD}, we briefly review the basic formulation of CD driving and introduce our inertial CD approach, explaining its
difference from trace-based variational AGP approaches. In Sec. \ref{ME}, we show how the ME is used to construct the inertial CD term and relate the nonlocal structure of the propagator and nested-commutator structure to the DLA. In Sec. \ref{Spinssys}, we discuss the numerical implementation and benchmark the approach in spin systems. In Sec. \ref{Crit}, we extend the analysis to quantum systems exhibiting spectral collapse and critical behavior. Finally, in Sec. \ref{Diss}, we summarize our conclusions and discuss future directions.

\section{Approximate CD frame}
\label{APPCD}

CD driving \cite{Demirplak2005,Berry_2009,jarzynski2013} is a quantum control protocol within the STA family that suppresses diabatic transitions, enabling the system to follow the adiabatic trajectory over finite evolution times. The CD term can be derived starting from the time-dependent Schr\"odinger equation
\begin{equation}
    i\hbar \partial_t |\Psi(t)\rangle = H_0(\boldsymbol{\lambda}(t)) |\Psi(t)\rangle ,
\end{equation}
with $\boldsymbol{\lambda} (t) $ being the set of time-dependent control parameters of the system Hamiltonian $H_0$. CD control rewrites the dynamics into a moving frame $|\tilde \Psi(t)\rangle = U^\dagger(\boldsymbol{\lambda}(t)) |\Psi(t)\rangle$, where the rotated states remain in the instantaneous eigenbasis trajectory. To simplify the notation, explicit time dependence will be omitted whenever possible. In this new basis, the Schr\"odinger equation becomes 
\begin{equation}
    i\hbar \partial_t |\tilde{\Psi}(t)\rangle = \left[ U^\dagger H_0 U + i \hbar\, \dot{\boldsymbol{\lambda}} \cdot \nabla_\lambda[U^\dagger] \, U \right] |\tilde{\Psi}(t)\rangle,
\end{equation}
where the rotated Hamiltonian $\tilde{H}_0=U^\dagger H_0 U$ is diagonal by construction, and $\tilde{\mathcal{A}}_\lambda:=- i \hbar \nabla_\lambda [U^\dagger] U \, $ represents the adiabatic gauge potential (AGP) in the moving frame~\cite{jarzynski2013,Berry_2009}. When the evolution is sufficiently slow, i.e. $|\langle m |\dot{\boldsymbol{\lambda}} \cdot\tilde{\mathcal{A}}_\lambda| n \rangle|  \ll|\epsilon_m-\epsilon_n|$ ($m \neq n $), the off-diagonal inertial couplings become negligible, and the adiabatic dynamics is recovered. However, for faster evolutions, these inertial couplings can induce transitions between instantaneous eigenstates, and the adiabatic approximation breaks down. CD driving therefore introduces an auxiliary Hamiltonian $H(t) = H_0(t)+H_{\mathrm{CD}}(t)$, with $H_{\mathrm{CD}}(t)$ designed to exactly counteract the inertial coupling,
\begin{equation}
\label{CD_definition_1}
 U^\dagger H_{\mathrm{CD}} U  + i \hbar\dot{\boldsymbol{\lambda}} \cdot \nabla_\lambda [U^\dagger] \, U   = 0.
\end{equation}
Although the exact CD formalism provides a way to precisely suppress diabatic transitions, its general implementation is challenging because constructing $U$ requires full spectral knowledge of $H_0$. Furthermore, the exact CD Hamiltonian typically contains highly nonlocal interactions, which severely limits its experimental feasibility. 

To overcome these limitations, variational CD approaches approximate the AGP within a restricted and experimentally accessible operator manifold \cite{sels_polkovnikov_minimizing}. Defining the basis within a predefined and accessible operator set, the CD is expressed in terms of this basis. First, we define the CD driving in terms of the AGP
\begin{equation}
    H_{\mathrm{CD}}(t) = \dot{\lambda}(t)\mathcal{A}_\lambda,
\end{equation}
where \(\mathcal A_\lambda=U\tilde{\mathcal A}_\lambda U^\dagger\) is the AGP in the original frame. For simplicity, we henceforth consider a single control parameter $ \lambda $, such that  $\nabla_\lambda \rightarrow \partial_\lambda$. By differentiating the rotated Hamiltonian $\tilde{H}_0$ and using $[\partial_\lambda \tilde{H}_0,\tilde{H}_0]=0$ \cite{jarzynski2013,sels_polkovnikov_minimizing,SW_DRAG},  the AGP satisfies 
\begin{equation}
\label{AGP_matrix}
    i\hbar \left( \partial_\lambda H_0 + \mathcal{M}_\lambda \right) = \left[ \mathcal{A}_\lambda, H_0 \right],
\end{equation}
where
\begin{equation}
    \mathcal{M}_\lambda =-\sum_n\partial_\lambda \epsilon_n({\lambda})    |n({\lambda})\rangle \langle n({\lambda})|
\end{equation}
is a generalized force operator. From the equations above, we can define the variational functional
\begin{equation}
\label{G_lam_action}
    G_\lambda[\chi] = \partial_\lambda H_0 + \frac{i}{\hbar} \left[\chi,H_0\right],
\end{equation}
where $\chi(t)$ denotes a variational approximation of the AGP. For the exact AGP, $\chi=\mathcal{A}_\lambda$, Eq.~(\ref{AGP_matrix}) ensures that the functional reduces to minus the generalized force operator, $G_\lambda[\mathcal{A}_\lambda]=-\mathcal{M}_\lambda$. Therefore, the closeness between an approximate $\chi$ and the exact AGP can be quantified through the Hilbert--Schmidt distance
\begin{equation}
   D[\chi] \equiv \mathrm{Tr}\left[ \left( G_\lambda+\mathcal{M}_\lambda\right)^2 \right]=\mathrm{Tr}\left[  G_\lambda^2\right]-\mathrm{Tr}\left[ \mathcal{M}_\lambda^2 \right].
\end{equation}
The second equality follows from the cyclic property of the trace and the fact that $\mathcal{M}_\lambda$ is diagonal in the eigenbasis of $H_0$. Since the generalized force $\mathcal{M}_\lambda$ is independent of $\chi$, the second term can be understood as a global shift, and therefore the best approximation of the AGP in the restricted manifold can be obtained by minimizing the functional
\begin{equation}
\label{action}
   \left.\frac{\delta S}{\delta \chi}\right|_{\chi=\chi^*} = 0 , \qquad S = \Tr[G_\lambda^2],
\end{equation}
usually referred to as the AGP action minimization, with $\chi^*$ denoting the operator that minimizes the action.

\begin{figure}[t]
    \centering
\includegraphics[width=\linewidth]{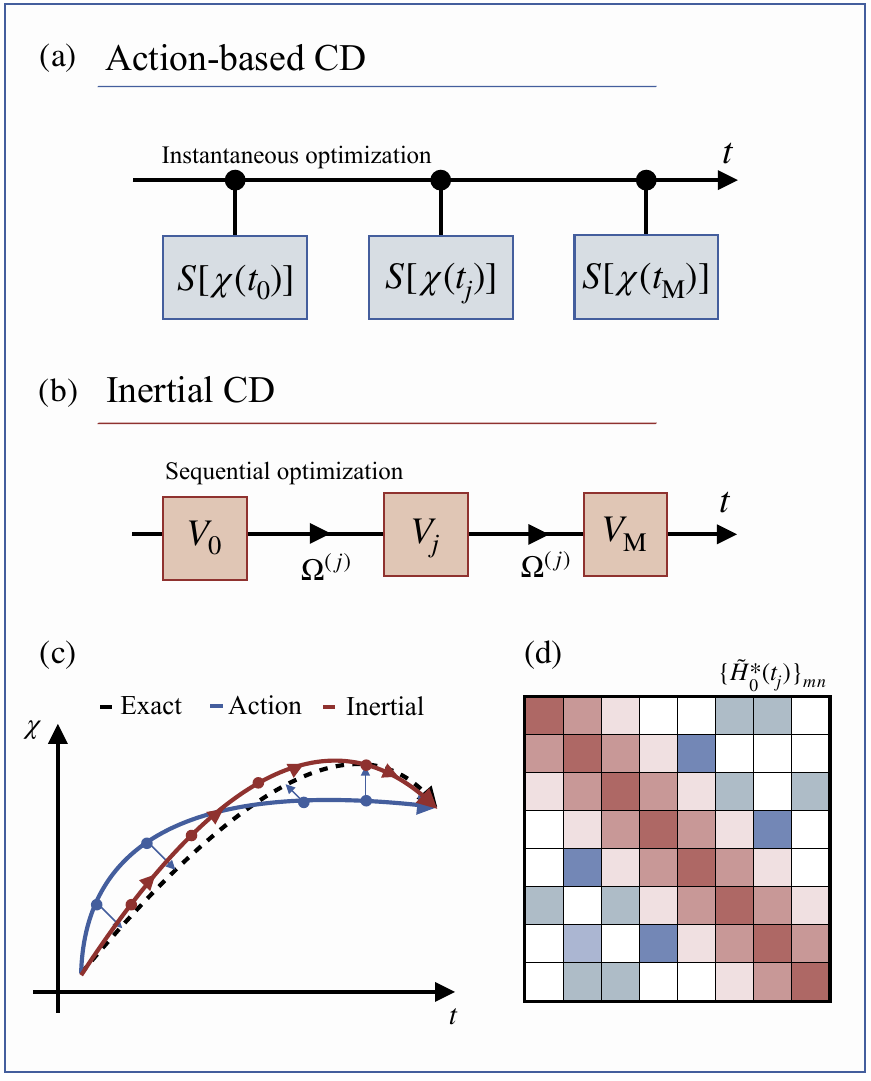}
\caption{Schematic comparison between action-based and inertial CD. (a) Action-based variational CD optimization scheme, where the structure of the approximate AGP  is determined independently at each time $t_j$ by minimizing the action $S[\chi(t_j)]$ [see Eq.~(\ref{action})]. (b) Inertial CD optimization scheme, where the moving frame is constructed sequentially through the Magnus exponent $\Omega^{(j)}$ over the $j$th time interval, so that each optimization step acts on the frame accumulated during the preceding evolution. The controls are chosen to minimize the off-diagonal part of the rotated Hamiltonian [see Eq.~(\ref{diagonalization_cost_function})], so that the constructed frame approximates the adiabatic moving frame. (c) Schematic time dependence of the control coefficients. Action-based CD (blue) approximates the exact AGP (dashed) pointwise, whereas inertial CD (red) selects the coefficients that suppress the accumulated diabatic error at each step. (d) Schematic structure of the rotated Hamiltonian $\tilde H_0^*(t_j)$ for a restricted ansatz. Red cells show the approximately diagonal Hamiltonian obtained in the frame generated by inertial CD, and blue cells the corresponding Hamiltonian obtained with action-based CD.}
    \label{fig: main}
\end{figure}
In this work, we introduce a novel variational protocol in which the CD Hamiltonian, rather than being obtained from the action-based variational formulation of Eq.~(\ref{action}), is constructed directly from the propagator formalism (see Fig. \ref{fig: main}). By solving the differential equation of Eq.~(\ref{CD_definition_1}), we can define an approximate adiabatic moving frame
\begin{equation}
\label{propagator_moving}
    V
    (t) = \mathcal{T} \exp\left( -\frac{i}{\hbar}\int_0^t \dot{\lambda}(\tau)\chi(\tau) \, d\tau\right)V(0),
\end{equation}
where $\mathcal{T}$ denotes the time-ordered operator and $V(0)$ is the transformation that diagonalizes the initial Hamiltonian $H_0(0)$. For \(\chi=\mathcal A_\lambda\), \(V(t)\) reproduces the exact eigenbasis transformation \(U(t)\), up to the chosen gauge, and \(V^\dagger(t)H_0(t)V(t)\) is therefore diagonal. The variational controls can thus be determined by minimizing the off-diagonal components of the Hamiltonian in this moving frame. To quantify deviations from the perfect diagonalization of the Hamiltonian, we introduce the cost function 
\begin{equation}
\label{diagonalization_cost_function}
    C[\chi] = \frac{\left\| \tilde{H}_0^* - \mathrm{diag}(\tilde{H}_0^*) \right\|_{\mathrm{tr}}}{\left\| \tilde{H}_0^* \right\|_{\mathrm{tr}}},
\end{equation}
where $\tilde{H}_0^* = V^\dagger(t)\, H_0\, V(t)$ is the Hamiltonian in the rotating basis and \(\operatorname{diag}(\cdot)\) denotes the diagonal part in the fixed reference basis. This formulation recasts the problem as a basis optimization, making it conceptually related to superadiabatic expansions \cite{Superadiabatic_1,superadiabatic_2,Superadiabatic_3} and DRAG techniques \cite{Drag_1,Drag_2,Drag_3} (see Appendix \ref{appI: techniques_comparation}). Unlike the instantaneous action in Eq.~(\ref{action}), however, the present cost depends on the accumulated propagator \(V(t)\) and is therefore explicitly history dependent, introducing an effective nonlocality in time into the variational optimization. The optimization thus targets the suppression of inertial couplings directly in the moving frame rather than the approximation of the exact AGP. In the ideal case where the variational manifold contains the exact solution ($\chi = \mathcal{A}_\lambda$), both approaches recover an exact CD driving.
 
\section{Magnus Expansion}\label{ME}
The success of the inertial CD approach relies on an accurate construction of the moving frame $V(t)$. Its evolution is governed by
\begin{equation}
\label{differential_equation}
    \partial_t V(t) = A(t)\, V(t),
\end{equation}
where $A(t) = -i\, \dot{\lambda}(t)\, \chi(t)=-i H_{\mathrm{CD}}(t)$ in natural units $(\hbar\equiv1)$. In the general non-commutative case, $[A(t_1), A(t_2)] \neq 0$, Eq.~(\ref{differential_equation}) does not admit a closed-form analytic solution, and it must be treated either numerically or through controlled approximations \cite{magnus_sliced,magnus_dalgaard_integrator}. Throughout this work, we employ the ME \cite{magnus_og,blanes_magnus} to approximate the unitary evolution operator. The ME provides a systematic representation of the evolution operator in terms of nested commutators of $A(t)$, capturing non-commutative effects order by order. Using the ME, the solution of Eq. (\ref{differential_equation}) can be written as
\begin{equation}
\label{ME_exponential}
   V(t) = \exp{\left(\Omega(t)\right)}V_0,
\end{equation}
where $V_0$ is the initial condition and $\Omega(t)$ is defined as
\begin{equation}
\label{Magnus_exp_formula}
    \dot{\Omega} = \sum_{k\geq 0}\frac{B _k}{k!}ad_\Omega^k(A(t)),
\end{equation}
being $B_k$ the Bernoulli numbers and $ad_\Omega(A)=[\Omega,A]$ is the adjoint operator. The convergence of the ME for bounded operators is ensured as long as $\int_{t_{j-1}}^{t_j}||A(t)||_2dt< \pi$ \cite{blanes_magnus}. 

For exact CD driving, the moving-frame transformation $U(t)$ diagonalizes $H_0(t)$ at every instant along the adiabatic trajectory, while the CD Hamiltonian cancels the associated inertial coupling. In our variational formulation, the inertial cancellation holds by construction, whereas the diagonality of the rotated Hamiltonian is imposed approximately through sequential optimization on a time grid $\{t_j\}$, with the Magnus exponent $\Omega^{(j)}=\Omega^{(j)}_1+\Omega^{(j)}_2+\cdots$ defined on each interval $[t_j,t_{j+1}]$. Here, $\Omega^{(j)}_k$ denotes the $k$th-order Magnus contribution evaluated over the $j$th interval. The moving-frame transformation is then written iteratively as \cite{magnus_iterated,magnus_sliced}
\begin{equation}
\label{step_ME}
    V_{j+1} = e^{\Omega^{(j)}} V_j. 
\end{equation}
The iterated ME naturally recasts the problem as a stepwise optimization problem, while the subdivision into finite time intervals allows the Magnus convergence condition to be enforced locally at each step.

We define the approximate CD Hamiltonian $H^*_{\mathrm{CD}}$ as 
\begin{equation}
\label{controled_AGP}
     H^*_{\mathrm{CD}}(t) = \dot{\lambda}(t)\sum_p x_p(t)\, \chi_p,
\end{equation}
where $x_p(t)$ are the variational parameters and \(\chi_p\in\mathcal H_{\rm CD}\) are the control operators spanning the CD operator manifold. The first two contributions to the Magnus exponent over the \(j\)-th time interval are
\begin{align}
\label{explicit_magnus}
\Omega^{(j)}_1 &= -  i\int_{t_j}^{t_{j+1}} H^*_{\mathrm{CD}}(t_1)\, dt_1, \\[6pt]
\Omega^{(j)}_2&= - \frac{1}{2} \int_{t_j}^{t_{j+1}} dt_1  \int_{t_j}^{t_{1}} dt_2 \left[ H^*_{\mathrm{CD}}(t_1), H^*_{\mathrm{CD}}(t_2) \right].
\end{align}
While the first-order term is restricted to the span of the native control operators, higher-order Magnus terms generate additional directions through nested commutators, thereby extending the effective operator space available to the diagonalization procedure. Effective DLA dynamics have previously been employed to generate virtual controls beyond the native operator manifold, notably in Floquet engineering \cite{oscilating_floq}, CD driving \cite{claeys_floq,petiziol,7wnd-t4xy}, and quantum algorithms \cite{QAOA_love,hatomura2026universal}. In contrast, we promote these effective directions from a Hamiltonian-engineering resource to an optimization resource, allowing the variational procedure to access the full expressivity of the DLA generated by the ansatz.  To characterize the structure of the accessible operator space, we employ a Cartan decomposition of the DLA $\mathfrak{g}$ generated by the total Hamiltonian \cite{petiziol,dAlessandro}. This decomposition takes the form $\mathfrak{g} = \mathfrak{h} \oplus \mathfrak{p}$, with the commutation relations
\begin{equation}
    [\mathfrak{h}, \mathfrak{h}] \subseteq \mathfrak{h}, \quad
    [\mathfrak{h}, \mathfrak{p}] \subseteq \mathfrak{p}, \quad
    [\mathfrak{p}, \mathfrak{p}] \subseteq \mathfrak{h}.
\end{equation}
By identifying $\mathrm{span}\{ H_0 \} \subseteq \mathfrak{p}$, and consequently $\partial_\lambda H_0 \in \mathfrak{p}$, the Cartan commutation relations together with Eq. (\ref{G_lam_action}) show that the CD generators live within $\mathfrak{h}$. Since \(\mathfrak h\) is closed under commutation, the ME preserves this algebraic structure, with \(\Omega_k\in\mathfrak h\) at every order. Importantly, higher-order Magnus terms can generate directions within \(\mathfrak h\) that are not explicitly contained in the native variational \(\mathrm{span}\{\chi_p\}\), thereby enlarging the effective operator space available in the optimization.


\section{Implementation in spin systems}\label{Spinssys}

In this section, we investigate the performance of the inertial CD protocol in spin systems. We first consider a two-qubit model, where the structure of the CD correction can be analyzed explicitly, and subsequently generalize to a many-body Ising chain. This progression allows us to assess the protocol from an analytically tractable setting to many-body systems.

Before discussing specific models, we briefly summarize the optimization procedure. The inertial CD protocol is optimized by suppressing the off-diagonal components of the rotated Hamiltonian [Eq.~(\ref{diagonalization_cost_function})] over the $\mathrm{M}$ time steps used to construct the discretized Magnus propagator [see Eq.~(\ref{step_ME})]. However, the diagonalization cost alone does not constrain the magnitude of the control fields. As a consequence, the optimization may favor solutions with large control amplitudes, as commonly encountered in unconstrained optimal-control formulations such as GRAPE~\cite{Grape1,Grape2}. To regularize the problem and account for the energetic cost associated with the auxiliary driving, we augment the objective function with a quadratic penalty term, analogous to regularization terms commonly used in optimal-control problems~\cite{dAlessandro,PRXQuantum.2.030203},
\begin{equation}
\label{cost_function_energy}
    \mathcal{J}_j[\{x_p\}]= C_j[\{x_p\}] +\mu \sum_p \int_{t_j}^{t_{j+1}} dt \,|x_p(t)|^2,
\end{equation}
where $\mu$  is a penalty parameter that controls the trade-off between diagonalization efficiency and control resources  (see Appendix \ref{appII: hyperparameters dependence}).

\subsection{Two-qubit benchmark}

\begin{figure}
    \centering
    \includegraphics[width=\linewidth]{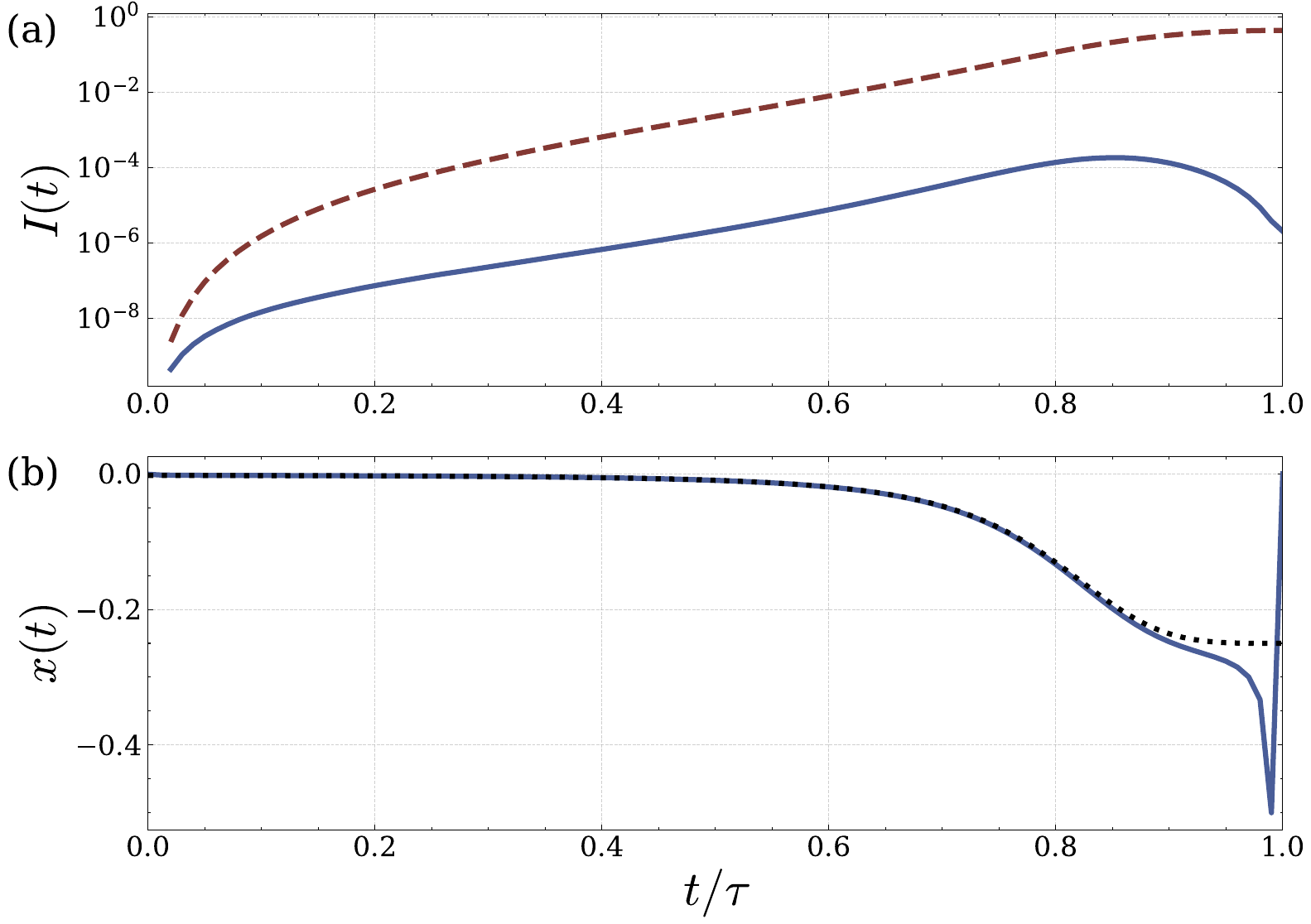}
    \caption{Two-qubit benchmark of the inertial CD protocol. (a) Instantaneous GS infidelity as a function of the normalized evolution time for adiabatic (red dashed) and inertial CD dynamics (blue solid).  (b) Exact action-based (black dotted) and inertial CD (blue solid) control coefficients. Parameters: $\tau=0.1$, $h_z=5$, $J=-1$, $\mu=10^{-3}$, and $M=100$.}
    \label{fig: two-qubit_cd}
\end{figure}

We first benchmark the inertial CD construction introduced in Eq.~(\ref{propagator_moving}) against the conventional action-minimization approach in a simple, analytically tractable two-qubit model. We consider the Hamiltonian
\begin{equation}
\label{eq:two_qubit_hamiltonian}
    H_0(t)
    =
    J \left(
        \sigma_z^1 \sigma_z^2
        +
        \sigma_x^1 \sigma_x^2
    \right)
    +
    h_z \left[\lambda(t)-1\right]
    \left(
        \sigma_z^1+\sigma_z^2
    \right),
\end{equation}
where $\sigma_\gamma^i$, with $\gamma=x,y,z$, denotes the Pauli operator acting on site $i$, $J$ is the spin--spin coupling, and $h_z$ is the longitudinal field. The scheduling function is defined as $\lambda(t)=\sin^2(\pi t/2\tau)$, where $\tau$ is the total evolution time. Despite its simple structure, the model provides a useful setting to assess the suppression of diabatic transitions along the evolution, since fast unassisted driving generates appreciable nonadiabatic excitations.  For this Hamiltonian, the exact AGP lies within the one-dimensional operator manifold
\begin{equation}
\label{eq:two_qubit_agp}
    \mathcal{A}_{\lambda}(t) = x(t) \left(\sigma_y^1\sigma_x^2 + \sigma_x^1 \sigma_y^2 \right).
\end{equation}
The coefficient $x(t)$ is obtained exactly by minimizing the action in Eq.~(\ref{action}), yielding
\begin{equation}
\label{analytic_two_qubit}
    x(t) = -\frac{Jh_z} {2(J^2+4(\lambda(t)-1)^2h_z^2)} .
\end{equation}
 Figure \ref{fig: two-qubit_cd}(a) compares the instantaneous infidelity, 
 \begin{equation}
 I(t) \equiv 1-\mathcal{F}(t)=1-|\langle \psi(t)|\psi_{GS}(t)\rangle|^2,
 \end{equation}
between the driven state $\ket{\psi(t)}$ and the instantaneous ground state (GS) $\ket{\psi_{GS}(t)}$ with and without the inertial CD correction. The inertial CD correction reduces the diabatic error by several orders of magnitude, keeping the driven state close to the instantaneous GS throughout the evolution. The exact-CD result is not shown, since in this case $I(t)$ vanishes identically, up to numerical precision.

Next, in Fig.~\ref{fig: two-qubit_cd}(b), we compare the exact control coefficient obtained from Eq.~(\ref{analytic_two_qubit}) with that obtained from the inertial optimization. Both controls exhibit a similar temporal profile over most of the evolution, with a deviation near the end of the ramp. Since the exact AGP is reachable within the variational ansatz and diagonalizes the rotated Hamiltonian by construction, both optimizations recover the exact AGP in the absence of regularization. The small discrepancy observed therefore results from the finite penalty $\mu$. However, the equivalence between action-based and inertial CD is not generic. In a many-body setting where the exact AGP lies outside the restricted variational manifold, the action and inertial formulations minimize distinct functionals and generally select different approximate controls.

\subsection{Ising chain}

We now move beyond the two-qubit benchmark and investigate the inertial CD protocol in a many-body spin chain. Specifically, we consider assisted adiabatic passage in an $N$-spin one-dimensional Ising model with transverse and longitudinal fields,\small
\begin{equation}
\label{Ising_equation}
    H_0(t) = (1-\lambda(t)) \sum_{i=1}^N h_x \sigma_x^i 
    - \lambda(t) \left(\sum_{i=1}^N h_z \sigma_z^i 
    + J \sum_{i=1}^{N-1}\sigma_z^i \sigma_z^{i+1}\right).
\end{equation}
\normalsize
Here, $J$ is the spin--spin coupling, and $h_x$, $h_z$ are the transverse and longitudinal fields, respectively.  The quantum Ising Hamiltonian constitutes a paradigmatic model for interacting many-body systems, with experimental realizations ranging from trapped ions \cite{Ising_beginners,RevModPhys.trap_ions} to cold atoms \cite{RevModPhys.rydberg_atoms,Rydberg_atoms_ising_2} and superconducting circuits~\cite{PhysRevB.80.052506,PhysRevB.82.024511}. It also provides a paradigmatic framework for exploring fundamental phenomena such as quantum phase transitions \cite{sachdev_qpt,Ising_PT_exper} and many-body correlations \cite{Ising_beginners,zhang2017observation}.
\begin{figure*}[t]
    \centering
\includegraphics[width=\linewidth,height=9cm]{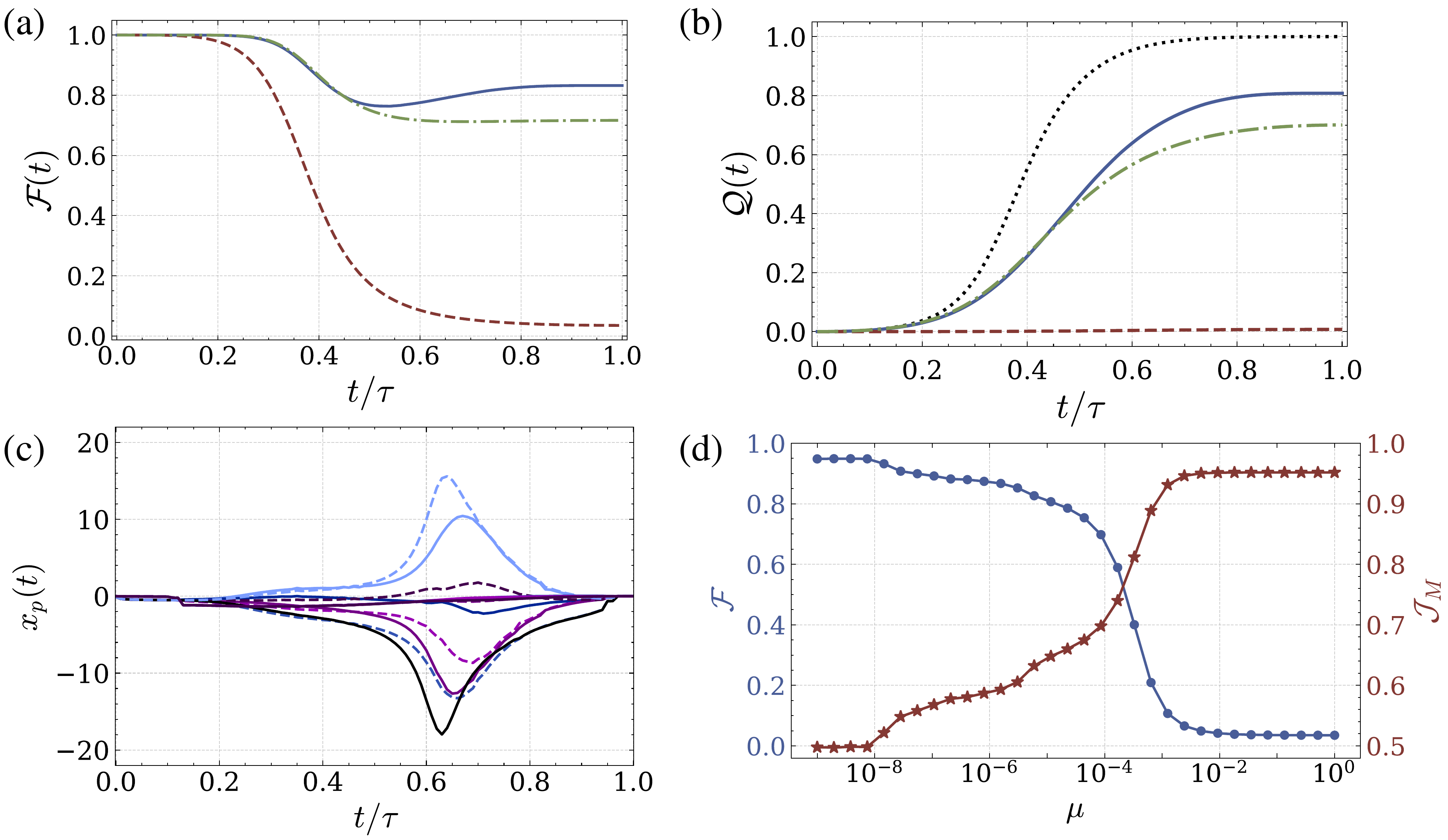}
  \caption{Quench dynamics of the longitudinal-field Ising model for $h_x=h_z=J=1/2$,  $N=5$, $M=100$, $\mu=5 \cdot 10^{-6}$, and $\tau=0.1\cdot 2\pi$. Results are shown for the unassisted adiabatic evolution (red dashed), the action-based variational AGP protocol (green  dot-dashed), and the inertial CD protocol (blue  solid). The control operators are obtained from the algebra spanned by the first-order Krylov expansion $(\ell = 1)$. (a) Instantaneous GS fidelity during the evolution. (b) Normalized ferromagnetic correlation density $\mathcal{Q}$, compared with its instantaneous-GS value (black dotted). (c) Optimized variational coefficients of the inertial CD protocol for $h_z=5 \cdot 10^{-2}$. (d) Final  diagonalization cost and final GS fidelity as functions of the regularization parameter $\mu$.}
    \label{fig:magnus_k1}
\end{figure*}

Unlike the two-qubit system, where the AGP lies within a closed, analytically tractable operator manifold, the many-body Ising model generates an increasingly complex operator structure. Exact AGP constructions therefore rapidly become intractable, motivating systematic approximation schemes. In this context, the Krylov expansion provides a natural variational basis for approximating the AGP~\cite{krylov_STA,wbbs-s8fs},
\begin{equation}
\label{krylov_nested}
    \chi(t) = i \sum_{\ell=1}^{L} x_\ell(t)\, \mathrm{ad}_{H_0}^{\,2\ell-1} \big( \partial_\lambda H_0(t) \big),
\end{equation}
where $x_\ell(t)$ are time-dependent variational parameters. This nested-commutator construction systematically enlarges the variational operator space without requiring spectral information. Increasing \(L\), however, generally generates operators of increasing body weight and spatial range, making their direct implementation progressively more demanding. We restrict the analysis to the first Krylov order, $L = 1$, which keeps the resulting operators relatively simple to implement. To increase the flexibility of the control and improve the state preparation, the terms obtained from the nested commutator are decomposed into independent operator components as
\begin{equation}
    \chi(t)=\sum_p x_p(t)\chi_p,
\end{equation}
where $x_p(t)$ are independent control amplitudes, and  $\chi_p$ denotes the elements of the operator pool obtained by decomposing $i[H_0,\partial_\lambda H_0]$ into its components.

\subsubsection{First-order ME}
We first analyze the protocol by retaining only the first-order Magnus contribution, thereby separating the effect of the inertial optimization from the additional commutator-generated directions arising at higher Magnus orders. We compare the GS fidelity during the evolution for unassisted driving, variational CD based on action minimization, and the inertial CD protocol, focusing on their respective convergence properties. Fig.~\ref{fig:magnus_k1}(a) shows the GS fidelity for a chain of $N=5$ spins under three protocols: unassisted evolution, CD control obtained via action minimization, and the inertial CD method. While both CD approaches significantly enhance the fidelity compared to the unassisted case, the inertial CD protocol achieves higher final fidelity. 

In Fig. \ref{fig:magnus_k1}(b), we illustrate the ferromagnetic correlation density \cite{schmitt2022quantum}

\begin{equation}
\mathcal{Q}(t) = \frac{1}{Q_0}\left| \sum_{i<i'}\langle \psi(t) |\sigma_z^i\sigma_z^{i'}| \psi(t) \rangle \right|,
\end{equation}
along the evolution. The ferromagnetic correlation captures the spin--spin correlations over all pairs normalized with $\mathcal{Q}_0 = N(N-1)/2$. For the final Hamiltonian considered, the ferromagnetic GS has $\mathcal{Q}=1$. The close agreement of \(\mathcal Q(t)\) with its instantaneous-GS value demonstrates that the inertial protocol not only achieves high state fidelity, but also accurately reproduces the behaviour of longitudinal many-body correlations characteristic of the target ferromagnetic GS~\cite{campo2014universality}. Fig.~\ref{fig:magnus_k1}(c) shows the optimized variational coefficients obtained with the inertial CD method. The coefficients exhibit pronounced peaks near the midpoint of the evolution, where the competing transverse-field and Ising-interaction energy scales, \((1-\lambda)h_x\) and \(\lambda J\), become comparable. In this region, the instantaneous GS changes from a transverse-field-dominated state to one dominated by longitudinal Ising correlations. This crossover is accompanied by a reduction of the instantaneous many-body gap. In the vicinity of the avoided crossing, nonadiabatic couplings increase as the energy gap decreases, requiring correspondingly stronger CD corrections. The observed amplitude profile reflects the sensitivity of the protocol to the reduction of the energy gap and how the penalty coefficient $\mu$ regularizes this response by penalizing excessively large variational coefficients. Fig. \ref{fig:magnus_k1}(d) illustrates the role of the penalty coefficient $\mu$ in the optimization. Increasing $\mu$ more strongly penalizes large variational coefficients, thereby reducing the available variational freedom and degrading the diagonalization efficiency. This demonstrates that the cost function provides a physically meaningful trade-off between diagonalization accuracy and control resources, allowing control constraints to be incorporated into the optimization. Therefore, reducing the penalty allows the optimization to reach fidelities close to unity, at the expense of larger variational coefficients.  Additional analysis of the optimization hyperparameters such as the number of time intervals $\mathrm{M}$ is provided in Appendix \ref{appII: hyperparameters dependence}. 

\begin{figure}[t]
    \centering
    \includegraphics[width=\linewidth]{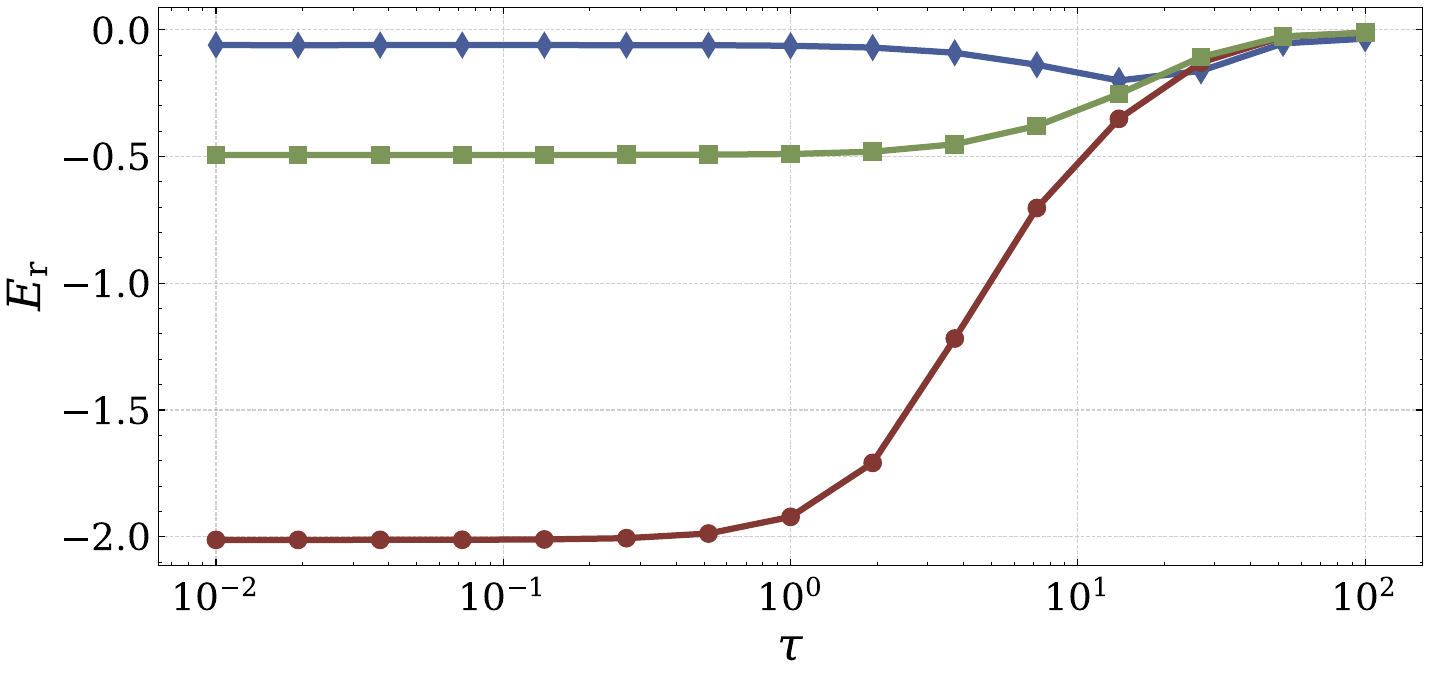}
  \caption{Final residual energy as a function of the total quench time \(\tau\). Results are shown for the inertial CD protocol (blue diamonds), the action-based CD protocol (green squares), and unassisted adiabatic evolution (red circles). The inertial protocol yields lower residual energies throughout the nonadiabatic regime considered. Parameters: $h_x = 1/2$, $J=1/2$,  $h_z=2 \cdot 10^{-3}$, $M=100$ and $\mu=10^{-6}$.}
    \label{fig: evolution_time_k1}
\end{figure}

Finally, in Fig.~\ref{fig: evolution_time_k1}, we report the final residual energy achieved by the different protocols at different evolution times. The residual energy quantifies the nonadiabatic excitations generated during the dynamics, and it is defined as 
\begin{equation}
    E_{\mathrm{r}}(\tau)=E_{\mathrm{GS}}(\tau)-\langle \psi(\tau)| H_0(\tau) |\psi(\tau)\rangle,
\end{equation}
where $E_{\mathrm{GS}}(\tau)$ denotes the final GS energy obtained by Hamiltonian diagonalization, and $\langle \psi(\tau)| H_0(\tau) |\psi(\tau)\rangle$ the energy of the driven state at final time $\tau$. In small-gap regimes, nonadiabatic couplings become increasingly difficult to suppress using restricted CD controls. In the gap-closing limit, the exact AGP, and hence the corresponding CD correction, can develop singular behavior. Although the GS fidelity directly quantifies the overlap with the target state, it does not fully characterize the broader dynamical behavior of the evolution. It is therefore useful to complement it with physical observables such as the longitudinal correlation density \(\mathcal Q\) and the residual energy \(E_{\mathrm r}\), providing not only a measure of state-preparation accuracy but also a broader picture of the evolution. Figure~\ref{fig: evolution_time_k1} shows that, for sufficiently long evolution times, all protocols approach the target GS and yield negligible residual energies, as expected. In contrast, for fast quenches, where diabatic excitations become significant, the inertial CD protocol consistently achieves lower residual energies than both the action-based CD and unassisted evolution, demonstrating more effective suppression of nonadiabatic excitations.

Overall, these results show that, already at the lowest-order Magnus approximation and within the same restricted control manifold, the inertial CD protocol surpasses conventional action-based variational AGP constructions for the fast quenches considered here. It more accurately eliminates the diabatic errors, keeping the evolved state closer to the instantaneous ground state, and thereby substantially reducing the residual energy. This result highlights the benefit of optimizing the accumulated moving-frame dynamics rather than solely minimizing an instantaneous AGP-based action. 

\subsubsection{Second-order ME}
\begin{figure*}[t]
    \centering
    \includegraphics[width=\linewidth]{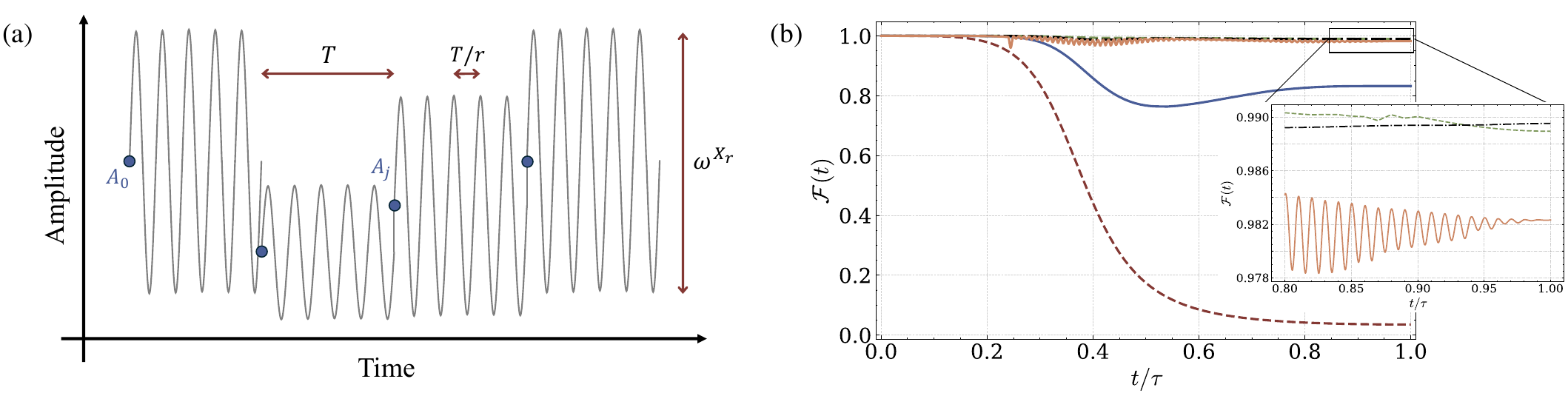}
  \caption{Second-order Magnus engineering in the longitudinal-field Ising model. (a) Schematic representation of the Fourier modulation used to enhance the effective second-order commutator contributions. (b) Instantaneous GS fidelity for the unassisted evolution (red dashed), the first-order inertial CD protocol with the $\ell=1$ control pool (blue solid), and the Fourier-enhanced second-order inertial CD protocol with the same $\ell=1$ control pool (orange solid). For comparison, results obtained with two explicitly enlarged operator spaces are also shown: the $\ell=2$ Krylov ansatz (dashed green) and the commutator-augmented  DLA ansatz (dashed-dotted black), both optimized using first-order ME. Parameters: $h_x=h_z=J=1/2$,  $N=5$, $M=100$, $\mu=5 \cdot 10^{-6}$, $\tau=0.1\cdot 2\pi$, and $\omega_0=1$.}
    \label{fig:magnus_k2}
\end{figure*}
We now extend the inertial CD protocol to second order in the ME, while retaining the restricted Krylov order \(\ell=1\). At first Magnus order, the effective generator remains confined in the manifold defined by the native control operators \(\{\chi_p\}\). By contrast, the second-order manifold also contains commutators \([\chi_p,\chi_q]\), providing access to additional directions within the DLA generated by the ansatz. However, the convergent construction of the ME causes the effective higher-order contributions to be increasingly suppressed as the time-slice duration decreases. To make these contributions appreciable while retaining the same native operator pool, we introduce a truncated Fourier parametrization of the variational coefficients, as illustrated in Fig.~\ref{fig:magnus_k2}(a).  Following the general idea of effective counterdiabatic driving, fast oscillatory controls can be used to generate otherwise inaccessible effective interactions through the Magnus expansion~\cite{petiziol,petiziolEPL}. We write
\begin{equation}
    x_p(t) = B_{p,0} + \sum_{r=1}^\mathcal{L} \left(\frac{\omega}{\omega_0}\right)^{X_r} \left[ A_{p,r} \sin(r \omega t) + B_{p,r} \cos(r \omega t) \right],
\end{equation}
where $\mathcal{L}$ is the number of harmonics, $\omega$ is the fundamental frequency, $\omega_0$ is a renormalization frequency, and $A_{p,r}$, $B_{p,r}$ are real amplitudes. The exponent
\begin{equation}
    X_r = 1 - \frac{1}{r+1}
\end{equation}
defines the frequency-dependent amplification profile of the Fourier harmonics. The corresponding $(\omega/\omega_0)^{X_r}$ scaling compensates for the high-frequency suppression of higher-order commutator contributions at the level of the effective generator. In the second-order construction below, we retain only the first harmonic, $\mathcal{L}=1$, for which we define the exponent $X_1=1/2$ to compensate for the suppression of the second-order Magnus term. The fundamental frequency $\omega=2\pi/T$ is set by the time-slice duration, and the coefficients $A_{p,r}$ and $B_{p,r}$ are taken to be constant within each interval. Substituting this expansion into the Magnus series (see Appendix~\ref{appIII: ME_fourier}), truncating the ME at second order, and retaining only the first Fourier harmonic, one obtains
\begin{subequations}
\label{k2_ME}
\begin{align}  
\Omega_1(T) &= -i \left(\frac{2\pi}{\omega}\right)\dot{\lambda}(t_j^*)\sum_p B_{p,0}\, \chi_p, \\[6pt]
\Omega_2(T) &= -\frac{\pi}{\omega \cdot \omega_0} \dot{\lambda}^2(t_j^*)  
\sum_{p< q} \left( A_{p,1} B_{q,1} - A_{q,1} B_{p,1} \right) [\chi_p, \chi_q].
\end{align}
\end{subequations}
Here $t_j^*$ denotes the midpoint of the $j$th interval. In Eq.~\eqref{k2_ME}, we see that, after the amplitude-profile engineering, the contributions of the first- and second-order terms share the same frequency dependence. This allows the effective commutator-generated directions to be introduced while the native control Hamiltonian remains restricted to the original operator pool.

In Fig.~\ref{fig:magnus_k2}(b), we analyze the optimization performance and fidelity convergence. We compare the inertial CD protocol implemented with first- and second-order ME using the same $\ell=1$ native ansatz pool. To assess the effective operator space accessed by the Fourier-enhanced protocol, we also include two extended constructions: the $\ell=2$ Krylov expansion and the commutator-augmented DLA ansatz both optimized using first-order ME. The latter is constructed by augmenting the $\ell=1$ operator space with the commutator-generated directions appearing at second Magnus order, 

\begin{subequations}
\begin{align}  
\mathfrak g^{(1)} &= \operatorname{span}\{\chi_p\},\\
\mathfrak g^{(2)} &= \operatorname{span} \left\{ [\chi_p,\chi_q]: \chi_p,\chi_q\in\mathfrak g^{(1)},\ p<q \right\},\\
\mathcal G^{(2)} &= \operatorname{span} \left( \mathfrak g^{(1)}\cup\mathfrak g^{(2)} \right).
\end{align}  
\end{subequations}

As expected, extending the operator basis through either the $\ell=2$ Krylov construction and to second-order DLA $\mathfrak{g}^{(2)}$ enlarges the accessible variational manifold and improves the suppression of diabatic transitions. Although these two constructions span different operator sets, they exhibit comparable performance for the present Ising benchmark, as both effectively introduce operators of similar many-body order. Remarkably, the Fourier-enhanced second-order inertial CD protocol achieves a comparable level of diabatic suppression while retaining only the native \(\ell=1\) control-operator pool. The additional directions are instead generated dynamically through the effective DLA, allowing the optimization to exploit an enlarged effective operator space without explicitly introducing the corresponding commutator operators as independent controls. In analogy with Floquet-based effective CD schemes~\cite{petiziol,petiziolEPL}, fast modulations therefore provide access to effective interactions that may be difficult to implement directly. Here, however, the commutator-generated directions are incorporated directly into the variational moving-frame optimization, rather than being engineered to reproduce a predetermined CD Hamiltonian.

\section{Dynamics across the finite-size critical region}\label{Crit}

To conclude, we investigate the performance of the inertial CD protocol in the vicinity of the gap closing of the quantum Ising model. The Ising model provides a paradigmatic setting to study how collective magnetic order emerges from microscopic spin interactions. Depending on the sign and relative strength of the spin-spin coupling and the external fields, the system can display paramagnetic  $|\rightarrow  \rightarrow\cdots \rightarrow\rangle$, ferromagnetic $|\uparrow \uparrow\cdots \uparrow\rangle/|\downarrow \downarrow\cdots \downarrow\rangle$, or antiferromagnetic order $|\uparrow \downarrow\cdots \uparrow \downarrow\rangle$. In the integrable Ising model ($h_z=0$),
\begin{equation}
\label{Integrable_Ising_equation}
H_I(t) = -h_x(t)\sum_{i=1}^{N}\sigma_x^i - J(t)\sum_{i=1}^{N-1}\sigma_z^i\sigma_z^{i+1},
\end{equation}
with $h_x,J>0$, the transverse field, which favors spin alignment along $x$, competes with the spin-spin interaction, which favors alignment along $z$. In the field-dominated regime, $h_x\gg J$, the GS is the paramagnetic state $|\rightarrow\rangle^{\otimes N}$, whereas in the interaction-dominated regime, $J\gg h_x$, the low-energy manifold is spanned by the two quasi-degenerate ferromagnetic configurations $|\uparrow\rangle^{\otimes N}$ and $|\downarrow\rangle^{\otimes N}$. Around $h_x/J=1$, the dominant term in the Hamiltonian switches from the transverse field to the spin-spin interaction, and the GS crosses over from paramagnetic to ferromagnetic order. In the thermodynamic limit $(N\rightarrow\infty)$, this crossover becomes a quantum phase transition (QPT) between the two phases.
At this critical point, the energy  gap closes and the relaxation time diverges.  For a finite-rate sweep, the dynamics therefore enters into the Kibble--Zurek impulse regime \cite{kibble1976topology,zurek1985cosmological,damski2005simplest,PRL_QPT_ZUREK}, in which the adiabatic approximation breaks down and the dynamics enters in a freeze-out region around the critical point. At the same time, the equilibrium correlation length grows near criticality, making the relevant diabatic fluctuations increasingly long-ranged. The extension of CD protocols to QPTs has been extensively studied \cite{campo2014universality,Ricardo_QPT_CD}, with particular emphasis on how the auxiliary controls suppress the enhanced diabatic excitations generated near criticality. For the Ising model \cite{Adolfo_QPT_ISING_PRL,visuri2026digitized}, it was shown that assisted transitionless driving across the critical point significantly reduces the density of excitations.  By employing the Jordan--Wigner transformation followed by Bogoliubov diagonalization (see Appendix \ref{appIV: exact_cd_ising}), they found that the CD Hamiltonian can be approximated as a series of truncated operators
\begin{equation}
\label{truncated_ising_cd}
    H^{[m]}_{\mathrm{CD}} \propto \sum_{i=1}^N \left(\sigma_z^i \cdots \sigma_x^{i+m-1}\sigma_y^{i+m} + \sigma_y^i \cdots \sigma_x^{i+m-1}\sigma_z^{i+m}\right),
\end{equation}
where $m$ denotes the truncation order. Far from the critical point, the dynamics is effectively local and low-order truncations ($m=1$) are sufficient to suppress excitations. However, near criticality, the relaxation time increases strongly as the excitation gap decreases. At the same time, the correlation length $\xi$, which characterizes the spatial extent of correlated fluctuations, also grows in the critical region. The relevant diabatic coupling, therefore, becomes increasingly nonlocal, and higher truncation orders $m$ are required to accurately approximate the CD correction. For the finite-size Ising model considered here, the thermodynamic QPT transition is replaced by a finite avoided crossing. The minimum gap therefore remains nonzero, but becomes progressively smaller as the system approaches the critical region. Correspondingly, increasingly long-range operator contributions are required in the CD Hamiltonian to suppress the enhanced diabatic transitions.

\begin{figure}[t]
    \centering
\includegraphics[width=\linewidth]{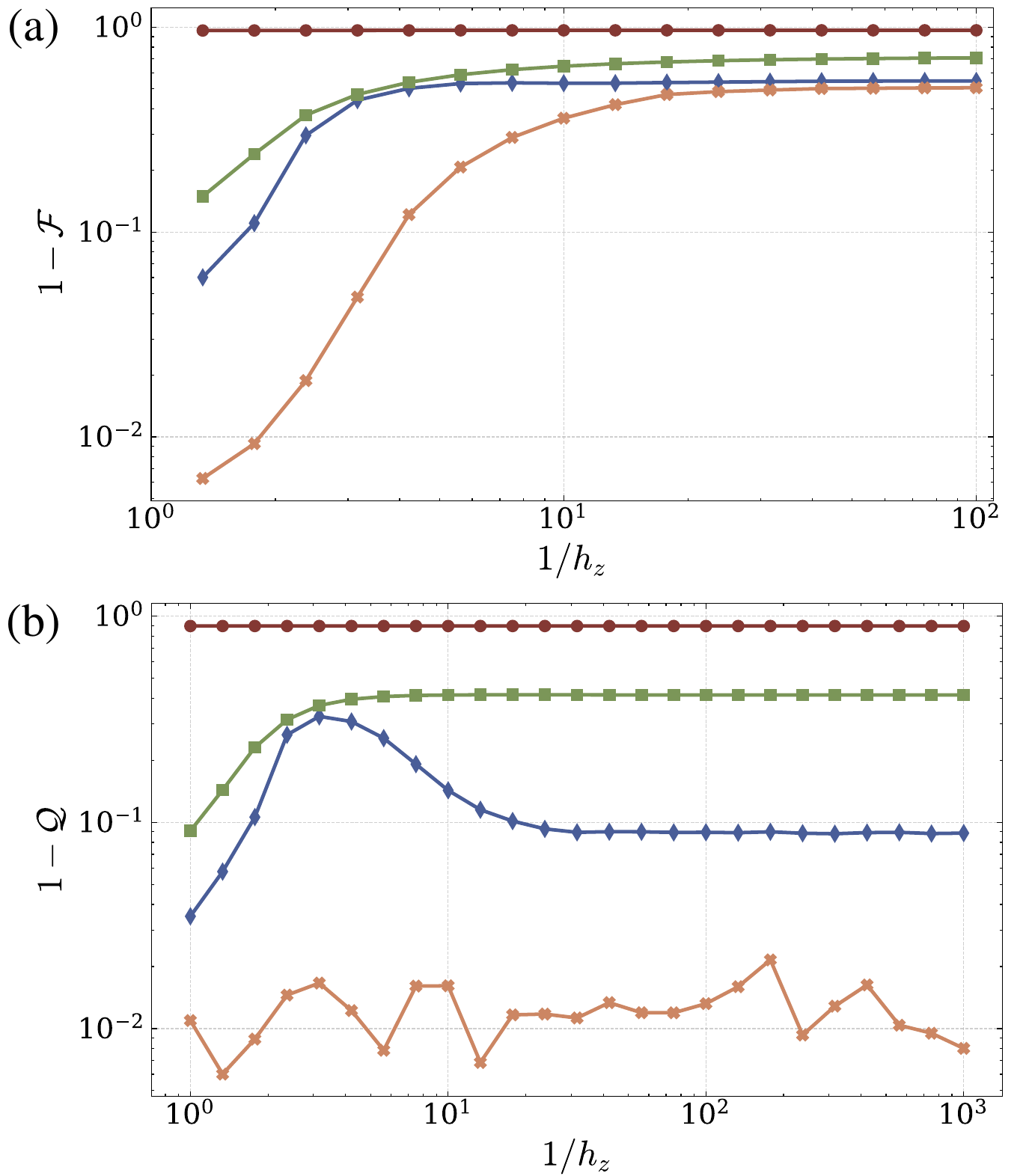}
    \caption{GS convergence for different energy gaps, set by different longitudinal fields $h_z$. (a) Final GS infidelity for the undriven trajectory (red circles), the action-based CD protocol (green squares), the first-order inertial CD protocol (blue diamonds), and the second-order inertial Magnus protocol (orange crosses) as a function of the longitudinal field $h_z$. (b) Deviation of the final ferromagnetic correlation $1-\mathcal{Q}$ for decreasing values of $h_z$. Parameters: $h_x=J=1/2$,  $N=5$, $M=100$, $\mu=5 \cdot 10^{-6}$, and $\tau=0.1\cdot 2\pi$.}
    \label{fig:QPT}
\end{figure}

In Figure \ref{fig:QPT}, we examine how the inertial CD behaves within a small-gap scenario, comparing our proposal against the action minimization. Starting from the quantum Ising model of Eq. (\ref{Ising_equation}), we simulate the GS dynamics for different values of the longitudinal field $h_z$. The longitudinal field explicitly breaks the $\mathbb{Z}_2$ symmetry and lifts the degeneracy between the two ferromagnetic configurations. As $h_z$ is reduced, the symmetry-breaking bias becomes weaker, and the lowest two energy levels approach one another, making the dynamics increasingly susceptible to diabatic transitions.

In Fig. \ref{fig:QPT}(a), we observe that the inertial CD protocol outperforms both the undriven and the action-minimized CD. At the second-order ME, the enlarged effective control manifold provides access to additional operator directions through the DLA $\mathcal G^{(2)}$, leading to a substantial improvement in the final-state preparation.
The first-order ME construction, although with a smaller margin, also outperforms the action minimization. Since both methods span the same instantaneous AGP operator manifold at this order, this improvement cannot be attributed to a richer control algebra. Instead, the result suggests that the inertial optimization procedure itself provides a more effective suppression of diabatic transitions.

In Fig.~\ref{fig:QPT}(b), we compare the deviation $1-\mathcal{Q}$ of the final magnetic correlation for the different protocols. Although the fidelity directly quantifies convergence to a particular GS, it may not fully capture the physically relevant behavior in the presence of a nearly degenerate low-energy manifold. As the longitudinal field decreases, the two lowest-energy states become increasingly close in energy \cite{Hastings2005}. In the limit $h_z\rightarrow 0$, the $\mathbb{Z}_2$ symmetry is restored and, for finite $N$, the final GS approaches the symmetric and antisymmetric superpositions of the two ferromagnetic configurations, $(|{\rm GHZ}_{\pm}\rangle=(|0\rangle^{\otimes N}\pm|1\rangle^{\otimes N})/{\sqrt{2}}$.
The magnetic correlation therefore provides a complementary measure of whether the control preserves the evolution within this low-energy ferromagnetic manifold. The results show that the second-order inertial CD protocol preserves the magnetic correlation to within approximately $1\%$, whereas, for the smallest gaps, action minimization induces appreciable leakage into higher-energy states.

Taken together, these results indicate that inertial CD provides an effective route for constructing approximate CD controls in the small-gap regime. At second order in the Magnus expansion, the Fourier parametrization enlarges the effective control manifold by generating additional higher-order operator contributions, thereby improving the convergence to the target GS. More notably, the first-order construction outperforms conventional action minimization despite spanning the same instantaneous AGP operator manifold. This comparison suggests that the improvement originates from the optimization procedure itself. In inertial CD, the controls are optimized iteratively over successive time intervals, with each step building on the evolution generated during the preceding intervals. This sequential construction retains information about the previously accumulated dynamics, introducing a temporal nonlocality into the optimization. Since this temporal nonlocality is absent from a purely instantaneous minimization of the AGP action, it may explain the improved performance of inertial CD in the small-gap regime. This robustness could be particularly relevant for larger systems, where the many-body level density grows exponentially with system size and the minimum gaps encountered along the evolution typically become smaller.

\section{Discussion and Conclusion}\label{Diss}

In this work, we introduced inertial CD, a variational protocol designed to suppress diabatic transitions through the construction of an approximate adiabatic moving frame. Rather than directly approximating the exact AGP, the method optimizes the AGP ansatz by constructing an inertial frame generated by the auxiliary control Hamiltonian and minimizing the off-diagonal diabatic errors. This provides a different variational perspective on AGP construction: instead of seeking the closest representation of the exact AGP within a restricted control manifold, inertial CD identifies the available control that most effectively suppresses nonadiabatic transitions. The optimal control therefore need not coincide with a truncated approximation of the exact CD term. When the accessible control manifold contains the exact CD operator, however, the method can recover transitionless dynamics that fully suppress the nonadiabatic couplings. The moving-frame transformation is constructed iteratively through a ME, so that each optimization step builds on the evolution accumulated over the preceding intervals.

We first benchmarked the method in a two-qubit model for which the exact CD term is known. In this setting, the inertial CD reproduces the relevant control structure and achieves high-fidelity state preparation. We then applied the protocol to the longitudinal-field Ising model using Krylov-inspired CD ansätze. Already at first order in the ME, the inertial construction outperforms the conventional action-based approximation obtained from the same instantaneous operator manifold.  We further showed that the method can be systematically improved by including higher-order Magnus contributions. In particular, fast Fourier modulations generate additional effective operators through the DLA, thereby increasing the expressivity of the control without explicitly enlarging the initial ansatz. This enhanced control manifold leads to improved performance in fast quenches and in the small-gap regimes of the finite-size Ising chain, where local CD approximations become progressively less effective, and the relevant diabatic dynamics acquire a collective character.

Overall, inertial CD driving provides a flexible route for constructing approximate CD controls in many-body systems. By combining moving-frame optimization and ME, the framework offers a systematic strategy for suppressing diabatic transitions in regimes where exact CD terms are nonlocal, experimentally demanding, or otherwise inaccessible.

\section*{Acknowledgment}
The authors would like to thank Pablo García-Azorín and M. Garc\'ia de Andoin for useful discussions and insightful comments. The project was partially supported from the Basque Government through the ELKARTEK program, project "KUBIBIT - Kuantikaren Berrikuntzarako Ikasketa Teknologikoa" (KK-2025/00079) and "BETISENS" (KK-2025/00107). X.C. acknowledges the project Grant No. PID2021-126273NB-I00, the Severo Ochoa Centres of Excellence program through Grant CEX2024-001445-S. This work was further supported by the European Union’s Horizon Europe programme through the QCFD project (Grant No. 101080085, HORIZON-CL4-2021-DIGITAL-EMERGING-02-10), OpenSuperQPlus100 (Grant No. 101113946, HORIZON-CL4-2022-QUANTUM-01-SGA); the ML4Q2 Cluster of Excellence (EXC 2004/2 – 390534769); the German Federal Ministry of Education and Research (BMBF) through the QSolid project (Grant No. 13N16149); and the Helmholtz Initiative and Networking Fund projects KA-QUS-02 (qFLOW) and KA-QUS-03 (QT-Batt).

\bibliography{bibl}
\onecolumngrid
\appendix


\section{Transitionless time-dependent techniques}
\label{appI: techniques_comparation}

Diabatic transitions constitute one of the main limitations in the coherent manipulation of quantum systems, and their suppression is central to the development of fast quantum control. A broad range of techniques has therefore been developed to reproduce, or approximate, adiabatic dynamics over finite time scales.

In this Appendix, we clarify the relation between the inertial CD protocol introduced in this work and several existing approaches. In particular, we compare inertial CD with variational CD driving based on action minimization, superadiabatic constructions, derivative removal by adiabatic gate (DRAG), and Magnus-based leakage-suppression protocols. Although these methods share the common goal of reducing unwanted transitions, they differ in the choice of frame, optimization target, and mechanism through which diabatic errors are suppressed.

\textbf{Variational CD methods} suppress diabatic excitations by approximating the exact AGP within a restricted operator manifold. The AGP generates the unitary transformation associated with adiabatic parameter changes. In natural units $(\hbar=1)$, its off-diagonal matrix elements are given by \cite{Berry_2009}
\begin{equation}
\label{appendix_exact_AGP}
    \langle m | \mathcal{A}_\lambda | n \rangle
    = i \langle m|\partial_\lambda n\rangle
    = -i \frac{\langle m | \partial_\lambda H | n \rangle}{E_m-E_n}.
\end{equation}
In the Hamiltonian eigenbasis, a Krylov approximation to the AGP takes the form
\begin{equation}
\label{appendix_krylov_AGP}
    \langle m | \chi | n \rangle
    = i \sum_{k=1}^{\ell} \alpha_k 
    \langle m|\mathrm{ad}_{H_0}^{2k-1}(\partial_\lambda H)|n\rangle
    = i \left[ \sum_{k=1}^{\ell} \alpha_k \omega_{mn}^{2k-1}\right]
    \langle m | \partial_\lambda H | n \rangle ,
\end{equation}
where $\omega_{mn}=E_m-E_n$. Comparing Eqs.~(\ref{appendix_exact_AGP}) and~(\ref{appendix_krylov_AGP}) shows that the Krylov ansatz approximates the inverse spectral dependence of the exact AGP through a polynomial in odd powers of the energy differences \cite{wbbs-s8fs}. In conventional variational CD, the coefficients of the ansatz are obtained by minimizing the distance between the variational operator and the exact AGP. This procedure does not require explicit knowledge of the exact AGP, and determines the best approximation within the chosen instantaneous operator manifold according to the AGP action [see Eq.~(\ref{action})]. Inertial CD follows a different variational principle.  Rather than minimizing the distance to the exact AGP, it constructs an approximate moving frame generated by the auxiliary control and optimizes the control coefficients by directly suppressing the off-diagonal components of the Hamiltonian in the rotated frame.

This difference in the variational construction leads to three main consequences. First, when the chosen AGP manifold is incomplete, action minimization selects the coefficients that best satisfy the instantaneous gauge-potential equation within that restricted ansatz, whereas inertial CD selects the coefficients that most effectively suppress the residual off-diagonal couplings in the constructed moving frame. The two optimization criteria therefore need not produce the same approximate CD Hamiltonian. Second, the Magnus construction naturally generates additional effective operator directions through nested commutators, allowing higher-order elements of the DLA to become accessible without explicitly including all of them in the bare AGP ansatz. Third, conventional variational AGP optimization is local in time, since the action is evaluated from the instantaneous Hamiltonian and its parameter derivative. In inertial CD, by contrast, the moving-frame propagator is constructed sequentially over finite time intervals, with each optimization step acting on the frame accumulated during the preceding evolution. The resulting control can therefore depend on both the accessible operator algebra and the previously accumulated dynamics.

\textbf{Superadiabaticity} suppresses diabatic errors through a sequence of iterated frame transformations \cite{Superadiabatic_3},
\begin{equation}
\label{superadiabatic_equation}
    H_j(t)=V_{j-1}^\dagger(t)H_{j-1}(t)V_{j-1}(t)    +i\dot{V}_{j-1}^\dagger(t)V_{j-1}(t).
\end{equation}
At each iteration, the transformed Hamiltonian is decomposed into a diagonal contribution $C_j \equiv V_{j-1}^\dagger H_{j-1}V_{j-1}$ and an inertial term $I_j \equiv i\dot{V}_{j-1}^\dagger V_{j-1}$. The objective is to construct a sequence of frames in which the relative magnitude of the inertial contribution becomes progressively smaller, quantified through an adiabatic quality factor
\begin{equation}
    Q_j=\int_0^{\tau}\frac{\|C_j\|}{\|I_j\|}\,dt.
\end{equation}
However, it is well known that higher orders in the superadiabatic expansion do not necessarily lead to a monotonic reduction of the residual inertial coupling, and beyond an optimal order the approximation may deteriorate. Inertial CD can therefore be viewed as a complementary frame-based route to diabatic suppression. Rather than relying solely on successive frame transformations to reduce the inertial term, the auxiliary CD control is used to construct a moving frame whose residual off-diagonal couplings are directly minimized. In this way, residual inertial effects are compensated variationally through the available control Hamiltonian, combining the moving-frame perspective of superadiabatic approaches with variational CD optimization.
 
\textbf{Derivative removal by adiabatic gate (DRAG)} was originally proposed to suppress leakage from the computational subspace during fast quantum gates. Its central idea is to add an auxiliary control quadrature, proportional to the time derivative of the primary control pulse, in order to cancel unwanted transitions between the computational and leakage levels. More generally, DRAG can be formulated within a perturbative frame-transformation framework, based on a time-dependent Schrieffer--Wolff (SW) transformation. We consider the Hamiltonian
\begin{equation}
\label{perturbative_leakage}
    H(t)=H_0(t)+V(t),
\end{equation}
where $H_0(t)$ generates the desired dynamics within the computational subspace and $V(t)$ contains the unwanted leakage couplings. By introducing a time-dependent anti-Hermitian generator $S(t)$, we define a dressed frame $e^{S}$ in which these couplings can be perturbatively suppressed and the leakage eliminated. The effective Hamiltonian in this frame can be written as
\begin{equation}
H_{\mathrm{eff}}=\sum_{n=0}^{\infty}\frac{1}{n!}\mathrm{ad}_S^n(H)+i\sum_{n=0}^{\infty}\frac{1}{(n+1)!}\mathrm{ad}_S^n(\dot S).
\end{equation}
And at leading order, the cancellation condition imposed by DRAG becomes
\begin{equation}
V+[S,H_0]+i\dot S=0.
\end{equation}
This structure can be connected to CD control by identifying the diabatic coupling as an effective leakage term. In this interpretation, DRAG/SW corrections iteratively suppress diabatic transitions by constructing a dressed frame in which the off-resonant couplings are perturbatively eliminated. If the exact adiabatic frame and the full AGP were available, this procedure would recover the exact CD correction. In the variational setting, however, the DRAG construction occupies an interesting middle position between superadiabaticity and inertial CD. Instead of suppressing by construction the inertial term or the off-diagonal terms in the rotated Hamiltonian, both the leakage and the inertial effects are removed order by order through the nested commutator expansion. This makes the method powerful in the perturbative regime, where the BCH series converges. However, the perturbative version fails near impulse regimes or small-gap regions, where diabatic couplings cannot be considered small, and SW needs to be replaced \cite{Drag_2,Drag_3}.

\textbf{Magnus-based control} provides a perturbative framework for suppressing leakage errors in driven quantum systems \cite{magnus_control_clerk}. Unlike DRAG, where unwanted couplings are removed through a time-dependent SW transformation, Magnus-based approaches construct an error propagator and cancel its contributions order by order in the ME. Starting from the decomposition in Eq.~(\ref{perturbative_leakage}), we add an auxiliary control Hamiltonian $W(t)$, so that the total Hamiltonian is $H_0(t)+V(t)+W(t)$, and write the evolution operator as
\begin{equation}
    U(t)=U_0(t)U_I(t),
\end{equation}
where $U_0(t)$, generated by $H_0(t)$, describes the target evolution and $U_I(t)$ contains the error dynamics. In the interaction frame defined by $U_0(t)$, the leakage and auxiliary control terms become
\begin{equation}
V_I(t)=U_0^\dagger(t)V(t)U_0(t),
\qquad
W_I(t)=U_0^\dagger(t)W(t)U_0(t),
\end{equation}
and the corresponding error propagator is
\begin{equation}
\label{error_leakage_clerk}
U_{\mathrm{err}}(\tau)=\mathcal{T}\exp\left[-i\int_0^\tau dt\,\big(V_I(t)+W_I(t)\big) \right].
\end{equation}
The auxiliary control $W(t)$ is designed such that, to a chosen order in the Magnus expansion, the error accumulated over the protocol duration vanishes within the computational subspace,
\begin{equation}
P_Q U_{\mathrm{err}}(\tau) P_Q
\simeq
\mathbb{I}_Q,
\end{equation}
where $P_Q$ projects onto the computational subspace.

The main distinction from CD driving lies in the control objective. Magnus-based leakage suppression generally aims to cancel the accumulated error over the complete evolution, whereas CD control seeks to suppress diabatic couplings along an adiabatic trajectory. For diabatic-error suppression, as in DRAG, the nonadiabatic coupling may be treated as the unwanted term in the corresponding moving frame. Once this coupling has been identified, Magnus-cancellation strategies can be applied. Inertial CD differs in that the ME is not used to cancel a predefined leakage Hamiltonian: it is used to construct the moving-frame propagator itself, whose residual off-diagonal components are then variationally minimized.

\section{Hyperparameter dependence} \label{appII: hyperparameters dependence}

\begin{figure}[h]
    \centering
    \includegraphics[width=\linewidth]{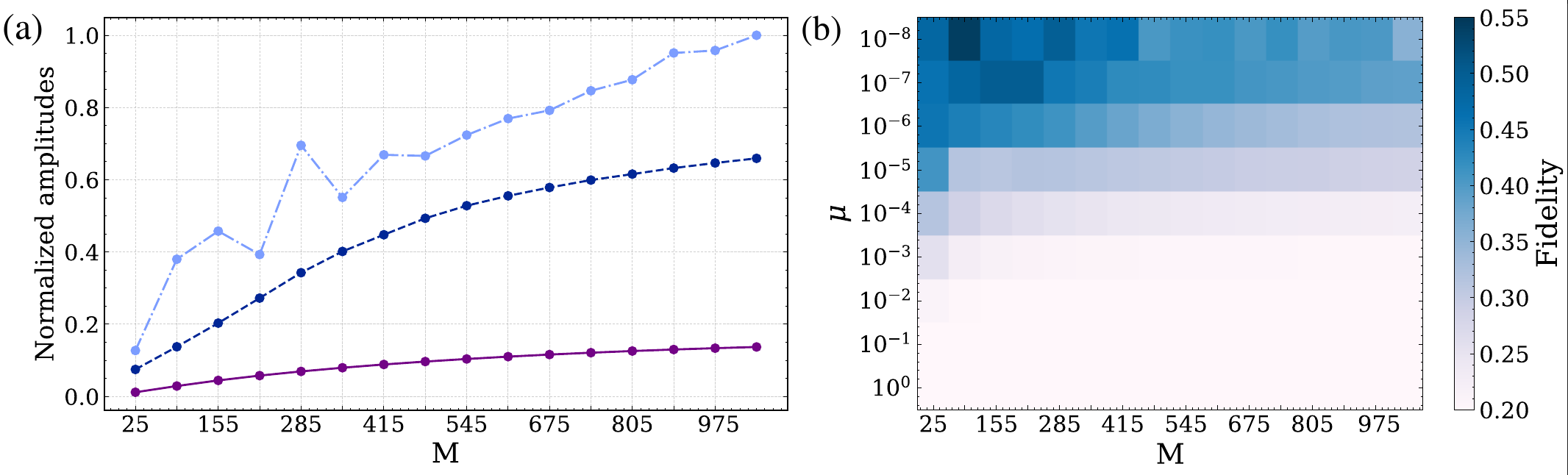}
  \caption{Behavior of the inertial CD optimization for different hyperparameter choices. (a) Normalized control amplitude as a function of the number of optimization intervals $\mathrm{M}$, shown for three penalty coefficients: $\mu_1=10^{-4}$ (solid), $\mu_2=10^{-6}$ (dashed), and $\mu_3=10^{-8}$ (dash-dotted). (b) Final GS fidelity as a function of the number of time slices $\mathrm{M}$, and different penalty coefficients. Parameters: $h_x=J=1/2$, $h_z=0.025 $, $N=5$,  and $\tau=0.1\cdot 2\pi$.  }
    \label{fig: mu_vs_nstep_vs_fid}
\end{figure}

In Sec.~\ref{Spinssys}, we extended the inertial CD diagonality cost function $C_j[{x_p}]$ by introducing an amplitude penalty, as defined in Eq.~\eqref{cost_function_energy}. The coefficient $\mu$ penalizes large control amplitudes, stabilizing the optimization and constraining the energetic cost of the auxiliary driving. As shown in Fig.~\ref{fig:magnus_k1}(d), smaller values of $\mu$ provide greater optimization freedom and generally lead to higher fidelities, at the expense of stronger control fields. The penalty coefficient $\mu$ must therefore be chosen to balance with diabatic-error suppression within experimentally feasible control amplitudes. In addition to $\mu$, the temporal discretization introduces a second relevant hyperparameter. In this Appendix, we analyze how the number of optimization intervals $\mathrm{M}$ affects the inertial CD protocol, and how it affects the control amplitudes.
Within the Magnus framework, the moving-frame propagator is constructed iteratively according to
\begin{equation*}
    V_{j+1} = \exp(\Omega^{(j)})\, V_j,
\end{equation*}
where $\Omega^{(j)}$ is evaluated over the interval $t\in[t_j,\,t_j+T]$, with $T=\tau/\mathrm{M}$ denoting the duration of each time slice. Increasing $\mathrm{M}$ provides a finer temporal resolution and can therefore improve the representation of the optimized trajectory. However, as $\mathrm{M}$ increases, the time slices get smaller, and the propagator approaches the identity, $\exp(\Omega^{(j)})\approx\mathbb{I}$. In this small-slice regime, the control applied within each interval produces a smaller rotation, so larger instantaneous amplitudes may be required to achieve a comparable dynamical effect. In Fig.~\ref{fig: mu_vs_nstep_vs_fid}(a), we report the normalized control amplitude for different penalty strengths $\mu$ and different numbers of time slices $\mathrm{M}$, quantified as
\begin{equation}
    \mathrm{Amplitude} = \frac{1}{x_{max}}\sum_p \sum_j \|x_p(t_j)\|,
\end{equation}
where $p$ labels the independent control parameters and $x_{max}$ is the maximum amplitude used for normalization. Within the parameter range considered here, the results show that decreasing the time-slice duration is accompanied by larger optimized control amplitudes. Thus, increasing the number of optimization intervals tends to require stronger controls.

In Fig.~\ref{fig: mu_vs_nstep_vs_fid}(b), we examine how the optimization hyperparameters affect the performance of the CD protocol by comparing the final GS fidelity for different penalty strengths and numbers of time slices. The penalty strength $\mu$ produces the dominant variation, with smaller penalties generally leading to higher fidelities. The number of slices $\mathrm{M}$ also affects the performance, although more weakly. Within the parameter range explored here, relatively coarse temporal discretizations achieve slightly better fidelity convergence than finer ones.

These results show that the performance of inertial CD depends on the hyperparameter selection. A careful choice of $\mu$ and $\mathrm{M}$ is therefore required to balance diabatic-error suppression, temporal resolution, and experimentally feasible control amplitudes.

\clearpage

\section{Truncated Fourier expansion in the ME } \label{appIII: ME_fourier}

\begin{figure}[b!]
    \centering
    \includegraphics[width=18cm,height=8cm]{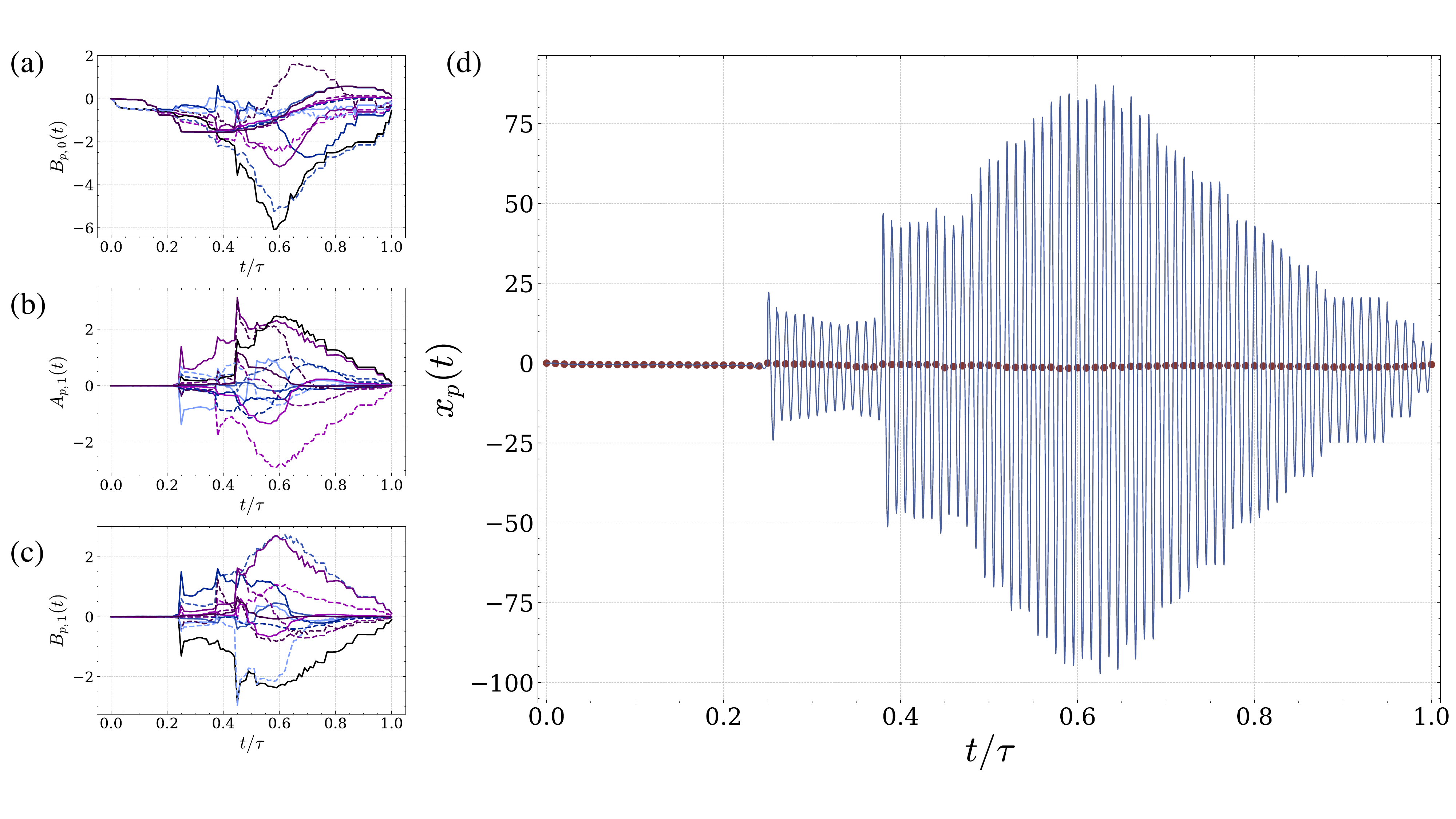}
    \caption{Variational control coefficients obtained with the inertial CD protocol at second order in the ME $(k=2)$. (a) Optimized zero-order, slowly varying coefficients $B_{p,0}$. (b) Optimized sine amplitudes $A_{p,1}$ of the first Fourier harmonic. (c) Optimized cosine amplitudes $B_{p,1}$ of the first Fourier harmonic. (d) Resulting time-dependent control amplitude reconstructed from the optimized Fourier components. The red markers denote the variational parameters optimized at each time interval, while the solid curve shows the corresponding Fourier-reconstructed control. Parameters: $h_x=J=1/2$,  $h_z=5 \cdot 10^{-2}$, $N=5$, $M=100$, $\omega_0 = 1$, $\mu=5 \cdot 10^{-6}$, and $\tau=0.1\cdot 2\pi$. }
    \label{fig: trace_vs_fidelity}
\end{figure}

In this work, we used the Magnus theory to define the inertial frame for the approximate CD control. Beyond providing a systematic approximation to the time-ordered propagator, the ME also generates additional effective control operators through its nested-commutator structure. As discussed in Sec. \ref{ME}, these higher-order contributions enlarge the DLA accessible to the control and can improve the GS convergence in the Ising chain. However, higher-order Magnus terms are strongly suppressed in the high-frequency regime, which limits their dynamical contribution unless the control amplitudes are scaled appropriately.

In this Appendix, we describe the truncated Fourier parametrization used to compensate for this suppression and to enhance the contribution of higher-order Magnus terms, thereby accessing a broader effective control manifold. First, let us define each control coefficient function by using a truncated Fourier expansion
\begin{equation}
    x_p(t) = B_{p,0} + \sum_{r=1}^\mathcal{L} \left( \frac{\omega}{\omega_0}\right)^{X_r} \left[ A_{p,r} \sin(r \omega t) + B_{p,r} \cos(r \omega t) \right].
\end{equation}
Here, $A_{p,r}$ and $B_{p,r}$ are the variational parameters optimized for each time grid $\{t_j\}$, $r$ labels the Fourier harmonic, and $\mathcal{L}$ denotes the truncation order. The frequency $\omega$ is chosen such that one oscillation period coincides with each optimization interval, $\omega={2\pi}/{T}$, with $T=t_{j+1}-t_j$, while $\omega_0$ defines a reference frequency. The factors $(\omega/\omega_0)^{X_r}$ rescale the amplitudes of the oscillatory components and compensate for the frequency suppression of higher-order Magnus contributions. In the following, we derive the resulting Magnus terms explicitly for the first-harmonic truncation $\mathcal{L}=1$.
 
For the first-order Magnus term, the oscillatory Fourier components average to zero over one complete driving period, so that only the slowly varying contribution survives. Assuming that $\dot{\lambda}(t)$ does not vary within one interval, the first-order generator reduces to
\begin{equation}
\Omega_1(T)
=
-i
\left(
\frac{2\pi}{\omega}
\right)
\dot{\lambda}(t_j^*)
\sum_p
B_{p,0}\chi_p,
\end{equation}
where $t_j^*$ denotes the midpoint of the time interval. For the second ME, the time-independent components vanish, and only the oscillatory terms survive. Restricting to the first-harmonic component, the control can be expressed as
\begin{equation}
x_p^{(1)}(t)
=
\left(
\frac{\omega}{\omega_0}
\right)^{X_1}
\left[
A_{p,1}\sin(\omega t)
+
B_{p,1}\cos(\omega t)
\right],
\end{equation}
the corresponding second-order Magnus term can be written as
\begin{equation}
    \Omega_2(T) = \frac{1}{2} \left(-i\right)^2\int_0^T dt_1 \int_0^{t_1} dt_2\, \dot{\lambda}(t_1)\dot{\lambda}(t_2) \sum_{p\neq q} \left[x_p^{(1)}(t_1) \chi_p, x_q^{(1)}(t_2) \chi_q \right],
\end{equation}
Since the terms with $p=q$ vanish. Assuming the fast-driving condition $\omega\gg|{\ddot{\lambda}}/{\dot{\lambda}}|$, we can treat $\dot{\lambda}(t_j^*)$ as constant over one period. Grouping each pair of noncommuting operators gives
\begin{equation}
   \Omega_2(T) = \frac{1}{2} \left(-i\right)^2\dot{\lambda}^2(t_j^*)\sum_{p< q} \int_0^T dt_1 \int_0^{t_1} dt_2 \{ x_p^{(1)}(t_1)x_q^{(1)}(t_2) - x_p^{(1)}(t_2)x_q^{(1)}(t_1)\} \left [ \chi_p,\chi_q \right].
\end{equation}

The double integral runs over the non-commuting operators, integrating over the oscillating controls. In these integrals, only the cross-terms of the Fourier expansion survive, and the integral of the coefficients reads as
\begin{equation}
    \int_0^Tdt_1\int_0^{t_1} dt_2 \sin(\omega t_1) \cos(\omega t_2) = \int_0^T dt_1\sin(\omega t_1) \left[ \frac{\sin(\omega t_1)}{\omega} \right]_0^{t_1} = \frac{T}{2\omega}.
\end{equation}
By substituting this integral, we get that the final effective control is
\begin{equation}
   \Omega_2(T) = -\frac{\pi}{\omega^2} \left( \frac{\omega}{\omega_0}\right)^{2X_r} \dot{\lambda}^2(t_j^*) 
\sum_{p< q} \left( A_{p,1} B_{q,1} - A_{q,1} B_{p,1} \right) [\chi_p, \chi_q]
\end{equation}
Where the exponent $X_r$ is the amplitude amplification, which prevents the ME from high-order decay. 

Finally, Fig. \ref{fig: trace_vs_fidelity} illustrates the oscillatory control structure obtained for the Ising-chain problem. Figs.\ref{fig: trace_vs_fidelity}(a)--(c) reveal the optimized parameters $B_{p,0}$, $A_{p,1}$ and $B_{p,1}$ associated with each control operator $\chi_p$, while Fig.~\ref{fig: trace_vs_fidelity}(d) displays the corresponding reconstructed oscillatory control profile. The key advantage of this parametrization is that the additional variational freedom is encoded in the periodic structure of the controls, allowing the second-order Magnus generator to access additional effective operator directions without increasing the number of time slices $\mathrm{M}$. Each additional harmonic introduces only two variational parameters per operator, $A_{p,r}$ and $B_{p,r}$, thereby enriching the effective control manifold without requiring additional intermediate time points.

\section{Exact CD in the Ising Model}\label{appIV: exact_cd_ising}

In this Appendix, we summarize the derivation of the exact CD Hamiltonian and its finite-range truncation [see Eq.~(\ref{truncated_ising_cd})] for the transverse-field Ising model \cite{Adolfo_QPT_ISING_PRL}. We define the Ising Hamiltonian as
\begin{equation}
    H_I = -\sum_{n=1}^N \left(\sigma_z^n \sigma_z^{n+1} + g\, \sigma_x^n \right).
\end{equation}
with periodic boundary conditions. This Hamiltonian corresponds to the integrable $h_z=0$ limit of the model studied in the main text. The longitudinal field, which explicitly breaks the $\mathbb{Z}_2$ symmetry and lifts the degeneracy between the two ferromagnetic configurations, is therefore absent in the following derivation.

The transverse-field Ising model can be mapped exactly onto a quadratic fermionic Hamiltonian through the Jordan--Wigner (JW) transformation,
\begin{subequations}
\label{JW_transformation}
\begin{align}  
\sigma_x^n &= 1 - 2 c_n^\dagger c_n, \\[4pt]
\sigma_z^n &= -(c_n + c_n^\dagger)\, \prod_{n'<n} (1 - 2 c_{n'}^\dagger c_{n'}),
\end{align}
\end{subequations}
where $c_n^\dagger$ and $c_n$ are fermionic creation and annihilation operators satisfying the anticommutation relations $\{c_n, c_{n'}^\dagger\} = \delta_{n,n'}$. This Hamiltonian conserves the $\mathbb{Z}_2$ parity $[H_I,P]=0$, with $P=\prod_{n=1}^{N}\sigma_x^n$. We restrict the analysis to the even-parity sector containing the GS. Applying a Fourier transform \begin{equation}
c_n=\frac{1}{\sqrt{N}}\sum_k c_k e^{ikn},
\end{equation}
followed by a Bogoliubov transformation \cite{PhysRev.127.1508}, the Hamiltonian can be simplified to
\begin{equation}
\label{diagonalize_ising}
    H_I = 2 \sum_{k>0} \psi_k^\dagger \left[ (g - \cos k)\, \sigma_x + \sin k\, \sigma_z \right] \psi_k.
\end{equation}
where $\psi_k^\dagger \equiv (c_k^\dagger, c_{-k})$ denotes the Nambu spinor, and $k = \pm \pi/N, \pm {3\pi}/{N},\dots, \pm (N-1)\pi/N$ are the odd pseudo-momenta. Here, each momentum mode is described by an independent effective two-level Hamiltonian, making the exact CD correction directly accessible. For a time-dependent transverse field $g(t)$, the CD Hamiltonian takes the form
\begin{equation}
H_{\mathrm{CD}} =-\dot{g}(t)\sum_{k>0}\frac{\sin k}{2\left(g^2+1-2g\cos k\right)}\,\psi_k^\dagger\sigma_y\psi_k .
\end{equation}
This Hamiltonian in momentum space, after transforming back to real space, generates CD operators with different spatial ranges. The exact CD Hamiltonian can therefore be decomposed \begin{equation}
H_{\mathrm{CD}}
=
\sum_m
\alpha_m(g,\dot g)\,
H_{\mathrm{CD}}^{[m]},
\end{equation}
where $m$ denotes the interaction range. For the spin convention used here,
\begin{equation}
H_{\mathrm{CD}}^{[m]}\propto\sum_{n=1}^{N}\left[\sigma_z^n\left(\prod_{j=n+1}^{n+m-1}\sigma_x^j\right)\sigma_y^{n+m}+\sigma_y^n\left(\prod_{j=n+1}^{n+m-1}\sigma_x^j\right)\sigma_z^{n+m}\right].
\end{equation}
Thus, increasing $m$ generates increasingly nonlocal strings of spin operators.

As shown in Ref.~\cite{Adolfo_QPT_ISING_PRL}, away from the critical point, the coefficients $\alpha_m$ decay exponentially with the interaction range $m$, so that the CD correction is effectively local and can be accurately approximated by truncating the expansion at finite range. As the critical point is approached, however, the characteristic decay length increases and the long-range terms become progressively more relevant. This provides the microscopic origin of the nonlocal CD corrections required near a QPT and explains why low-order local approximations of the AGP become progressively less accurate in the critical region.

\end{document}